\documentclass[aps,prb,10pt,twocolumn,amsmath,amssymb,superscriptaddress,showpacs,preprintnumbers,footinbib,longbibliography,11pt]{revtex4-1}
\usepackage{geometry}
\usepackage[T1]{fontenc}
\usepackage{textcomp}
\usepackage{txfonts}
\usepackage{bm}
\usepackage{natbib}
\usepackage{graphicx,color,epsfig}
\usepackage{xspace}
\DeclareFontFamily{OT1}{pzc}{}
\DeclareFontShape{OT1}{pzc}{m}{it}{<-> s * [1.2] pzcmi7t}{}
\DeclareMathAlphabet{\mathpzc}{OT1}{pzc}{m}{it}

\begin{document}

\title{Magnetic Vortex Crystals in Frustrated Mott Insulator}

\author{Y. Kamiya and C. D. Batista}
\affiliation{
  Theoretical Division, T-4 and CNLS, Los Alamos National Laboratory, 
  Los Alamos, New Mexico 87545, USA
}

\date{\today}

\begin{abstract}
  Quantum fluctuations become particularly relevant in highly frustrated quantum magnets and can lead to new states of matter. 
  We provide a simple and robust scenario for inducing magnetic vortex crystals in frustrated Mott insulators. By considering a quantum paramagnet that has a gapped spectrum with six-fold degenerate  low-energy modes, we study the magnetic-field-induced condensation of these modes. 
  We use a dilute gas approximation to demonstrate that a plethora of multi-$\mathbf{Q}$ condensates are stabilized for different combinations of exchange interactions. 
  This rich quantum phase diagram includes magnetic vortex crystals, which are further stabilized by symmetric exchange anisotropies. Because skyrmion and domain-wall crystals have already been predicted and  experimentally observed, this novel vortex phase completes the picture of emergent crystals of topologically nontrivial spin configurations.
\end{abstract}


\maketitle

\section{Introduction}
The emergence of topological spin textures in solids triggered an enormous interest because of their relevance for spin-electronic technology. Outstanding examples are the crystals of magnetic skyrmions that were recently discovered in noncentrosymmetric magnets with the B20 structure $MX$ ($M$ is a transition metal and $X=\text{Si, Ge}$)~\cite{Muhlbauer2009Skyrmion,Munzer2010,Yu2010Real-space}  and also in a Mott insulator Cu$_2$OSeO$_3$.\cite{Seki2012Observation,Seki2012Magnetoelectric,Onose2012Observation,Seki2012Formation,Adams2012Long-Wavelength,White2012Electric} 
A magnetic skyrmion is a hedgehog-like spin texture that wraps a sphere when mapped on the spin space. 
Crystals of these topological textures emerge in the above materials from competition between Dzyaloshinskii-Moriya and exchange interactions.\cite{Muhlbauer2009Skyrmion} Similar to Abrikosov vortex lattices in type-II superconductors, skyrmion crystals can be driven by injecting an electronic current in the metallic compounds.\cite{Jonietz2010Spin,Schulz2012Emergent} In contrast, Mott insulators allow for energetically more efficient manipulations of the skyrmion crystals because these spin textures induce a spatial modulation of electric dipole moments that can be driven by electric-field gradients.\cite{Seki2012Magnetoelectric,White2012Electric}

After this sequence of discoveries, it is natural to ask if crystals of topological spin configurations can emerge under more general conditions. While skyrmion crystals,\cite{Okubo2012Multiple} soliton crystals,\cite{Kamiya2012Formation,Selke1988ANNNI} and $Z_2$ vortex crystals~\cite{Rousochatzakis:arXiv1209.5895} have been predicted for classical spin systems, we are not aware of any prediction of crystals of usual (i.e., Abelian) magnetic vortices. In this paper we demonstrate that magnetic vortex crystals arise in a class of frustrated quantum magnets from a multi-$\mathbf{Q}$ Bose-Einstein condensate (BEC) of low-energy magnetic modes induced by the magnetic field. Because these crystals emerge from frustration, the intervortex distance is controlled by the ratio between competing exchange constants and can be tuned with external pressure.

The phenomenon of Bose-Einstein condensation appears in different realizations of bosonic gases. Atomic gases~\cite{Bloch2008Many-body} and superconductivity (condensation of Cooper pairs) are two prominent examples in which bosons normally condense into a zero-momentum single-particle state. 
Quantum magnets provide an alternative realization of Bose gases because spin operators of different ions commute with each other.\cite{Giamarchi2008bose} 
For instance, a lattice of $S=1/2$ moments can be exactly mapped into a gas of hardcore bosons on the same lattice.\cite{Matsubara1956Lattice} 
The notion of the Bose-Einstein condensation is only approximated in real spin systems because the total number of bosons (magnetization) is not strictly conserved. Anisotropic contributions arising from dipolar and spin-orbit interactions break the U($1$) symmetry of global spin rotations along an external magnetic field. 
Nevertheless, it has been established that  the description of the Bose-Einstein condensation offers a very good approximation when the anisotropy terms are weak.\cite{Giamarchi2008bose} The advantage of having weak anisotropy terms is that the boson density can be tuned  by applying a magnetic field, which works as a chemical potential. 
Another advantage of magnetic incarnations of Bose gases is that there are many materials in which the bosons condense at a nonzero-momentum single-particle state. 
This situation is quite common when the system has a highly frustrated exchange interaction because it can lead to a single-boson dispersion $\omega^{}_{\mathbf{k}}$ with minima at low-symmetry points of the Brillouin zone.
For $p$-fold symmetric lattices, for instance, such a strong frustration leads to degenerate minima of $\omega^{}_{\mathbf{k}}$ at different ${\bf Q}$ vectors related by $p$-fold rotations along the symmetry axis. 
Bosons can then condense in a single-${\bf Q}$ BEC state or in a linear combination of single-particle states with different ${\bf Q}$ vectors (multi-${\bf Q}$ BEC). As we will demonstrate here, this characteristic of frustrated quantum magnets opens the exciting possibility of stabilizing magnetic vortex crystals under rather general conditions.

\begin{figure}[b]
    \includegraphics[width=\hsize]{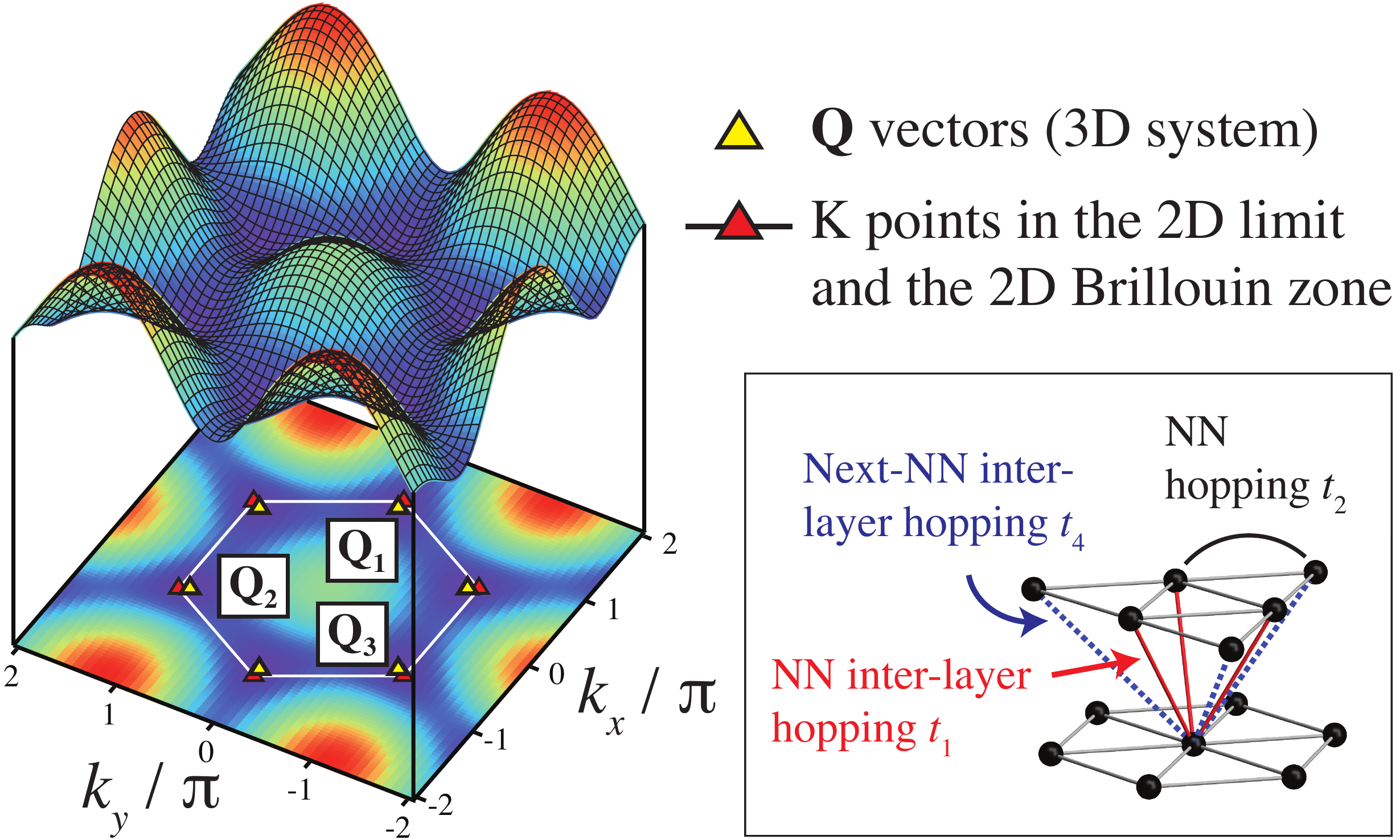}
  \caption{%
    \label{fig:dispersion}%
    Single-boson dispersion $\omega^{}_\mathbf{k}$ characterized by six-fold degenerate minima at low-symmetry positions $\pm \mathbf{Q}_{1 \le n \le 3}$ (shown for $k_z = 0$), which can be realized in the lattice shown in the inset yielding $\mathbf{Q}_1 = (2\alpha,\alpha,\alpha)$, $\mathbf{Q}_2 = (-\alpha,\alpha,0)$, and $\mathbf{Q}_3 = (-\alpha,-2\alpha,-\alpha)$.
    In Ba$_3$Mn$_2$O$_8$, $\alpha$ is slightly smaller than $2\pi/3$.\cite{Stone2008Singlet-Triplet}
  }
\end{figure}

Previous studies of multi-$\mathbf{Q}$ condensates in frustrated quantum spin systems, such as triangular lattice antiferromagnets~\cite{Nikuni1995Hexagonal,Veillette2005Ground,Deformed2011Griset} and helimagnets,\cite{Ueda2009Magnon} considered the minimal case where the system has only two different lowest-energy modes $\mathbf{k} = \pm \mathbf{Q}$. In this situation, the two possible condensates are a single-$\mathbf{Q}$ spiral state and a double-$\mathbf{Q}$ coplanar state, neither of which is a vortex crystal.
Actually, vortex crystals arise from $p$-$\mathbf{Q}$ condensates with $p \geq 3$. 
We will derive various multi-$\mathbf{Q}$ BEC solutions corresponding to vortex crystals by considering highly frustrated quantum spin systems with six-fold degenerate lowest-energy modes (Fig.~\ref{fig:dispersion}). 
Moreover, we will see that small anisotropy terms dominate interaction effects in the low-density limit, namely, close enough to the quantum critical point (QCP) that divides the magnetically ordered and the paramagnetic phases. Remarkably, this effect significantly enlarges the region where a particular type of vortex crystal is stabilized.

%

To illustrate our point, we consider a hardcore boson model on a lattice of  triangular layers stacked in a period-$3$ structure along the $c$ axis. The choice of this lattice is motivated by the highly frustrated quantum paramagnet Ba$_3$Mn$_2$O$_8$.\cite{Uchida2002High-field,Stone2008Singlet-Triplet,Samulon2008Ordered,Samulon2009Asymmetric,Samulon2010Anisotropic,Suh2011Nonuniversal} Ba$_3$Mn$_2$O$_8$ consists of triangular layers of weakly coupled antiferromagnetic $S=1$ spin dimers (Fig.~\ref{fig:BMO}). 
Each dimer is predominantly in the singlet state at low fields, and the lowest-energy excitation is a triplet state that propagates with well-defined momentum (triplon).\cite{Giamarchi1999coupled,Giamarchi2008bose} The triplon dispersion is gapped at low fields and has six-fold degenerate minima (Fig.~\ref{fig:dispersion}).
Because of the finite energy gap, triplons are only thermally activated below the critical magnetic field, $H=H_{c1}$, where the gap is closed. A BEC is stabilized at zero temperature ($T=0$) for $H \ge H_{c1}$,\cite{Giamarchi1999coupled,Nikuni2000BEC,Giamarchi2008bose} opening the possibility of multi-$\mathbf{Q}$ ordered states.
In an idealized situation without anisotropy, the model has U($1$) symmetry along the field direction and the type of BEC is determined by the effective triplon-triplon interactions. The arbitrarily low triplon concentration close enough to the QCP allows for a controlled and robust analytical approach by expanding in the small lattice-gas parameter.\cite{Beliaev1958Energy}
As mentioned above, we also discuss the effect of anisotropy that becomes relevant  close enough to the QCP.

\section{Model}

\begin{figure}[t]
  \includegraphics[width=0.92\hsize]{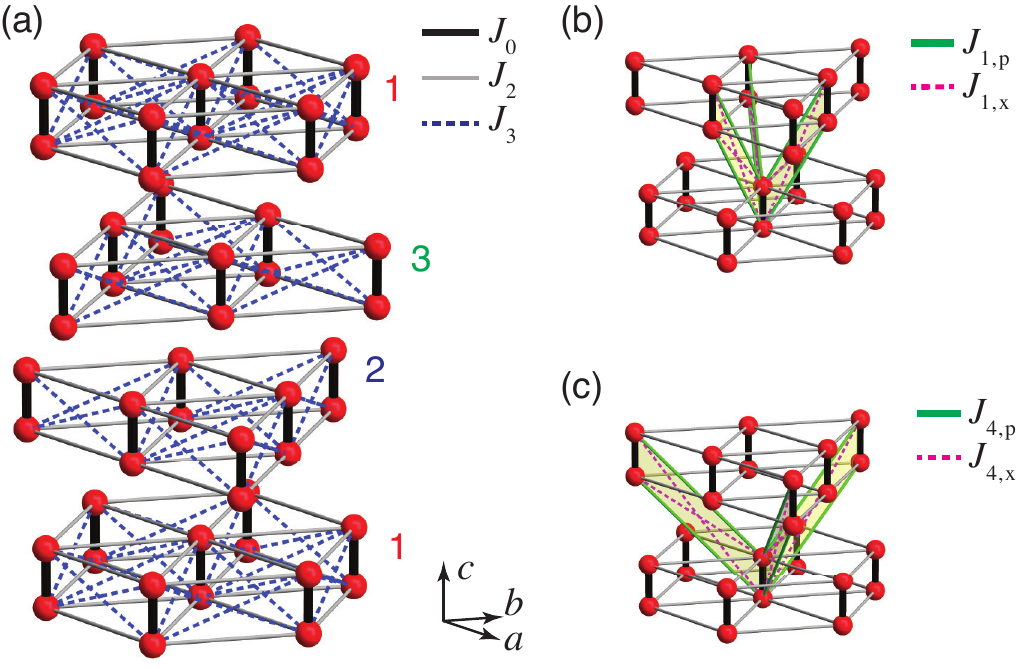}
  \caption{%
    \label{fig:BMO}%
    (a) Magnetic (Mn$^{5+}$) lattice of Ba$_3$Mn$_2$O$_8$.
    (b) and (c) NN and next-NN exchange couplings between adjacent bilayers.
  }
\end{figure}
%

The low-energy effective Hamiltonian for a spin-dimer antiferromagnet near the magnetic-field-induced QCP is
\begin{align}
  \mathpzc{H} & = \mathpzc{H}_{\,2} + \mathpzc{H}_{\,4},~~
  \mathpzc{H}_{\,2} = \sum_{\mathbf{k}} \omega_{\mathbf{k}}^{} \left(b^{\dagger}_\mathbf{k} b^{\;}_\mathbf{k} + \frac{1}{2}\right),
  \notag\\
  \mathpzc{H}_{\,4}^{} &= \frac{1}{2N} \sum_{\mathbf{k},\mathbf{k}',\mathbf{q}} \left(U + V_{\mathbf{q}}^{}\right) b^{\dagger}_{\mathbf{k}+\mathbf{q}} b^{\dagger}_{\mathbf{k}'-\mathbf{q}} b^{\;}_{\mathbf{k}'} b^{\;}_{\mathbf{k}}.
  \label{eq:ham}
\end{align}
Here, $b^{\dagger}_{\mathbf{k}} = N^{-1/2} \sum_{i=1}^{N} e^{i\mathbf{k}\cdot\mathbf{r}_i}\, b^{\dagger}_{i}$, 
and $b^{\dagger}_{i}$ ($b^{\;}_{i}$) is a bosonic creation (annihilation) operator of an $S^z = 1$ triplon in the dimer $i$ relative to the singlet background. Thus, we keep only two low-energy states of each dimer.
$U$ is the on-site hardcore potential to exclude unphysical states.\cite{Kotov1998Novel}
$V_{\mathbf{q}}$ is the Fourier transform of the microscopic off-site boson-boson interactions. 
We assume that the single-particle dispersion $\omega_{\mathbf{k}}^{}$ is characterized by six-fold degenerate minima as illustrated in Fig.~\ref{fig:dispersion}. For $H < H_{c1}$, the spectrum has an energy gap that can be controlled by the external magnetic field. 
For the moment, we will  exclude anisotropy terms that break the U($1$) symmetry of $\mathpzc{H}$.

The magnetic lattice system of Ba$_3$Mn$_2$O$_8$~\cite{Uchida2002High-field,Stone2008Singlet-Triplet,Samulon2008Ordered,Samulon2009Asymmetric,Samulon2010Anisotropic,Suh2011Nonuniversal} provides a perfect realization of $\mathpzc{H}$ (Fig.~\ref{fig:BMO}).
Inelastic neutron scattering measurements determined the spin Hamiltonian.\cite{Stone2008Singlet-Triplet} 
Each spin dimer is coupled by an antiferromagnetic exchange $J_{0} = 19.05(4)\mathrm{K}$, and spins on different dimers are coupled by $J_2$ and $J_3$ on the same layer and also by several other interlayer exchanges $J_{1,\mathrm{p}}$, $J_{1,\mathrm{x}}$, $J_{4,\mathrm{p}}$, and $J_{4,\mathrm{x}}$ [Figs.~\ref{fig:BMO}(a)--\ref{fig:BMO}(c)], which are much smaller than $J_{0}$.\footnote{%
  A slightly different set of interlayer exchange couplings was assumed in Ref.~\onlinecite{Stone2008Singlet-Triplet}. However, the exchange couplings described in the text also reproduce the measured triplon dispersion. The reason for such a redundancy is that each triplon hopping intensity between dimers is not related to a unique exchange coupling constant but is a function of several different exchange coupling constants between the associated spins. 
}

Because of these interdimer exchange interactions, a triplon can propagate in the 3D lattice. The microscopic hopping process includes the intralayer hopping, $t_2 \propto J_2 - J_3$, between nearest-neighbor (NN) dimers as well as $t_1 \propto J_{1,\text{p}} - J_{1,\text{x}}$ and $t_4 \propto J_{4,\text{p}} - J_{4,\text{x}}$, respectively, between NN and next-NN dimers on adjacent bilayers.\footnote{%
  $t_1 = -0.91(1)\mathrm{K}$, $t_2 = 1.76(1)\mathrm{K}$, and $t_4 = -0.29(1)\mathrm{K}$ were estimated for Ba$_3$Mn$_2$O$_8$ by the inelastic neutron scattering in Ref.~\onlinecite{Stone2008Singlet-Triplet}.
} 
Without these interlayer hopping processes, the minima of $\omega_{\mathbf{k}}^{}$ would be located at the K points on the 2D Brillouin zone edge. The finite values of $t_1$ and $t_4$ shift the $\mathbf{Q}$ vectors from such commensurate high-symmetry points  to incommensurate wave vectors, $\mathbf{k} = \pm \mathbf{Q}_n$ ($1 \le n \le 3$),  within the same reciprocal plane (see Fig.~\ref{fig:dispersion}).\footnote{%
  For $t_4 = 0$, two lines of minima appear in $\omega_{\mathbf{k}}^{}$ (Ref.~\onlinecite{Rastelli1988Competition}). This accidental degeneracy is lifted when $t_4 \ne 0$ is included.
}
Because these $\mathbf{Q}$ vectors are not connected by reciprocal unit vectors,  the lowest-energy single-triplon excitation becomes six-fold degenerate.
In addition to the on-site hardcore potential $U$, triplons are subjected to off-site density-density interactions $V_1 \propto J_{1,\text{p}} + J_{1,\text{x}}$, $V_2 \propto J_2 + J_3$, and $V_4 \propto J_{4,\text{p}} + J_{4,\text{x}}$ when they occupy adjacent dimers connected by the hopping paths of $t_1$, $t_2$, and $t_4$, respectively.

\section{Instability analysis of the field-induced Bose-Einstein condensation: Dilute Bose gas approximation}

\subsection{Ground state energy}
The triplon excitation spectrum becomes gapless at $H = H_{c1}$, signalling an instability towards formation of a  BEC  at the six-fold degenerate single-particle states $\mathbf{k} = \pm \mathbf{Q}_{1 \le n \le 3}$.
This instability is associated with a divergent transverse spin susceptibility at these $\mathbf{Q}$ vectors, i.e., 
the BEC state  corresponds to a magnetically ordered state for spin components perpendicular to the external field. 
The order parameter comprises the $\mathbf{k} = \pm \mathbf{Q}_{1 \le n \le 3}$ Fourier components of the transverse magnetization.
Therefore, to predict the spin structure of the ordered phase we need to determine the condensate distribution among $\mathbf{k} = \pm\mathbf{Q}_{1 \le n \le 3}$.
Quantum fluctuations provide the selection mechanism for interacting systems.~\cite{Nikuni1995Hexagonal,Veillette2005Ground,Ueda2009Magnon,Deformed2011Griset,Jackeli2004Frustrated}
Because the boson density vanishes at the field-induced QCP, the relevant effective interaction can be computed very accurately by using Beliaev's low-density approximation.\cite{Beliaev1958Energy} Through minimization of the resulting effective Hamiltonian, we can determine the BEC state right above $H = H_{c1}$ in a  reliable way.~\cite{Nikuni1995Hexagonal,Veillette2005Ground,Ueda2009Magnon,Deformed2011Griset,Jackeli2004Frustrated}

\begin{figure}
  \includegraphics[width=\hsize]{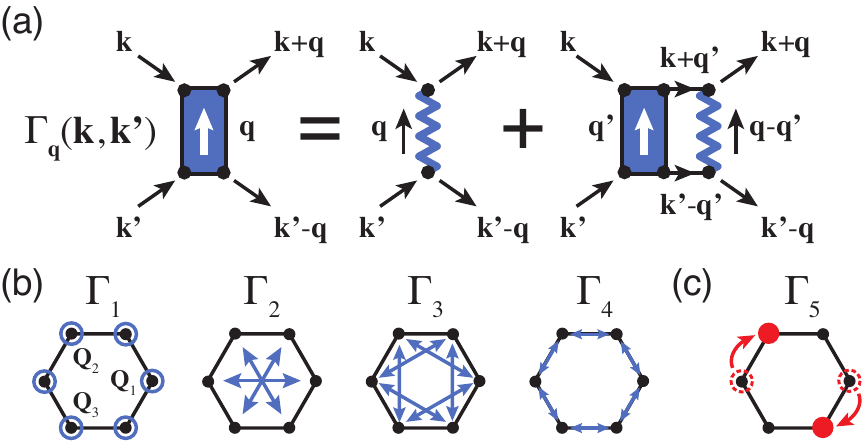}
  \caption{%
      \label{fig:vertex}%
      (a) Effective vertex from ladder diagrams. The filled square (wavy line) represents the effective (bare) potential.
      (b) Diagonal and (c) off-diagonal effective interactions in the GL theory.
  }
\end{figure}

We calculate the effective interaction in the long-wavelength limit, ${\bf k} \simeq \pm \mathbf{Q}_n$, by adding the ladder diagrams shown in Fig.~\ref{fig:vertex}(a).
The calculation is performed in the static limit (i.e., for zero total frequency) because  we are only interested in the ground state.
The interaction vertex, $\Gamma_{\mathbf{q}}(\mathbf{k},\mathbf{k}')$, for incoming triplons with momenta $\mathbf{k}$ and $\mathbf{k}'$ and momentum transfer $\mathbf{q}$ is asymptotically exact in the dilute limit $H \simeq H_{c1}$ (see Appendix~\ref{app:ladder} for details of this calculation).\cite{Beliaev1958Energy}
Once we obtain the interaction vertices, we can write down the Ginzburg-Landau (GL) expansion of the ground-state energy density, $E_{\text{eff}}$, with asymptotically exact GL expansion coefficients.
For deriving this GL theory, we only need to take into account the $\mathbf{k} = \pm \mathbf{Q}_{1 \le n \le 3}$ modes that have a divergent susceptibility.
Formally, such a GL expansion is obtained by replacing the bare interaction vertices, $U + V_\mathbf{q}$, involving $\mathbf{k},\mathbf{k}' \in \{\pm\mathbf{Q}_{1 \le n \le 3}\}$ with $\Gamma_{\mathbf{q}}(\mathbf{k},\mathbf{k}')$. Then, we approximate the condensates by $N^{-1/2} \langle b^{}_{\pm\mathbf{Q}_n} \rangle \equiv \sqrt{\rho_{\pm\mathbf{Q}_n}^{\;}} \exp({i\phi_{\pm\mathbf{Q}_n}})$ by using the standard Bogoliubov prescription.
We obtain
\begin{align}
  E_{\text{eff}}
  &=
  \,-\mu \sum_{n=1}^{3} \left(
  \rho_{\mathbf{Q}_{n}}^{\;} + \rho_{-{\mathbf{Q}}_{n}}^{\;}
  \right)
  + \frac{\Gamma_1}{2} \sum_{n=1}^{3} \left(
  \rho_{\mathbf{Q}_{n}}^2 + \rho_{-{\mathbf{Q}}_{n}}^2
  \right)
  \notag\\[-2pt]
  &+ \Gamma_{2} \sum_{n=1}^{3} 
  \rho_{\mathbf{Q}_{n}}^{\;} \rho_{-{\mathbf{Q}}_{n}}^{\;}
  + \Gamma_{3} \sum_{n < m} 
  \left(
  \rho_{\mathbf{Q}_{n}}^{\;} \rho_{\mathbf{Q}_{m}}^{\;}
  + \rho_{-{\mathbf{Q}}_{n}}^{\;} \rho_{-{\mathbf{Q}}_{m}}^{\;}
  \right)
  \notag\\[1pt]
  &+ \Gamma_{4} \sum_{n < m}
  \left(
  \rho_{\mathbf{Q}_{n}}^{\;} \rho_{-{\mathbf{Q}}_{m}}^{\;}
  + \rho_{-{\mathbf{Q}}_{n}}^{\;} \rho_{\mathbf{Q}_{m}}^{\;}
  \right)
  \notag\\[1pt]
  &+ 2{\Gamma_{5}}\sum_{n < m}
  \sqrt{
  \rho_{\mathbf{Q}_{n}}^{\;} \rho_{-{\mathbf{Q}}_{n}}^{\;}
  \rho_{\mathbf{Q}_{m}}^{\;} \rho_{-{\mathbf{Q}}_{m}}^{\;}
  }
  \cos\left(\Phi_n - \Phi_m\right),
  \label{eq:GL}
\end{align}
where $\mu = g\mu_{\text{B}}^{}(H - H_{c1})$ and 
\begin{align}
  \Phi_n = \phi^{}_{\mathbf{Q}_n} + \phi^{}_{-{\mathbf{Q}}_n}.
  \label{eq:Phi}
\end{align}
Equation~\eqref{eq:GL} is universal as long as the minima of $\omega_{\mathbf{k}}$ are six-fold degenerate at incommensurate wave vectors. 
The GL coefficients are given by the following vertices:
\begin{align}
  \Gamma_{1} 
  & = \Gamma_{0}\left(\mathbf{Q}_n, \mathbf{Q}_n\right),~~
  \notag\\
  \Gamma_{2} 
  &= \Gamma_{0}\left(\mathbf{Q}_n, -\mathbf{Q}_n\right) + \Gamma_{-2\mathbf{Q}_n}\left(\mathbf{Q}_n, -\mathbf{Q}_n\right),
  \notag\\
  \Gamma_{3} 
  & = \Gamma_{0}\left(\mathbf{Q}_n, \mathbf{Q}_m\right) + \Gamma_{\mathbf{Q}_m - \mathbf{Q}_n}\left(\mathbf{Q}_n, \mathbf{Q}_m\right),
  \notag\\
  \Gamma_{4} 
  & = \Gamma_{0}\left(\mathbf{Q}_n, -\mathbf{Q}_m\right) + \Gamma_{-\mathbf{Q}_m - \mathbf{Q}_n}\left(\mathbf{Q}_n, -\mathbf{Q}_m\right),
  \notag\\
  \Gamma_{5} 
  & = \Gamma_{\mathbf{Q}_m - \mathbf{Q}_n}\left(\mathbf{Q}_n, -\mathbf{Q}_n\right) 
  + \Gamma_{-\mathbf{Q}_m - \mathbf{Q}_n}\left(\mathbf{Q}_n, -\mathbf{Q}_n\right),
  \label{eq:gamma}
\end{align}
where $1 \le m \ne n \le 3$. 
As illustrated in Fig.~\ref{fig:vertex}(b), $\Gamma_{1 \le \nu \le 4}$ represents the effective density-density interaction between the condensate triplons. 
The $\Gamma_5$ vertex represents a process in which a pair of triplons with momenta $\mathbf{k} = \pm\mathbf{Q}_n$ is annihilated and  a different pair is created with momenta $\mathbf{k}' = \pm \mathbf{Q}_{m \ne n}$ [Fig.~\ref{fig:vertex}(c)].
Because of its off-diagonal nature, the $\Gamma_5$ term is the only one that depends on the relative phases  $\Phi_n-\Phi_m$.

\begin{figure}[b]
  \includegraphics[width=\hsize]{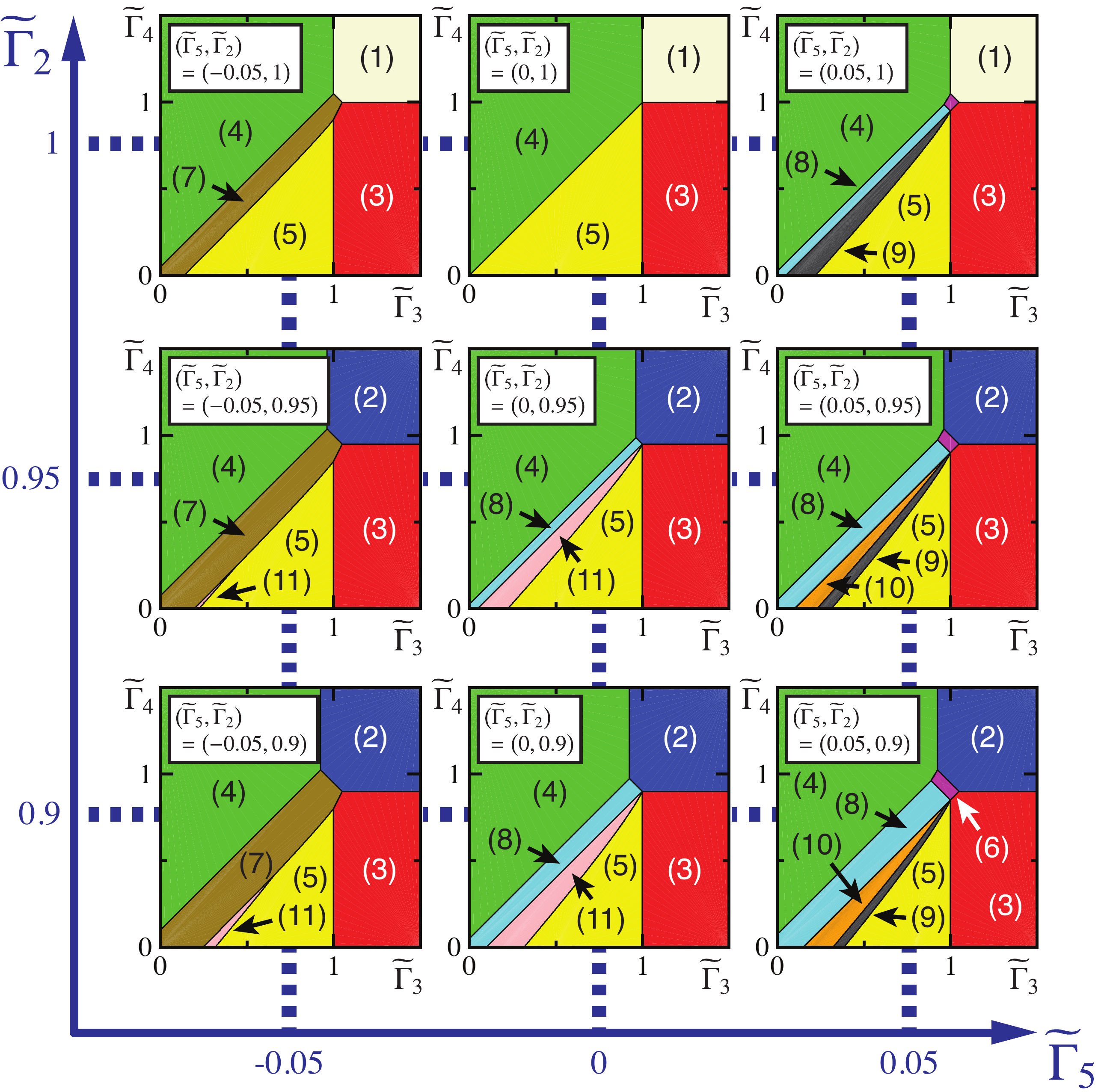}
  \caption{%
    \label{fig:phase-diagram1}%
    Phase diagram as a function of $\widetilde{\Gamma}_\nu \equiv \Gamma_\nu / \Gamma_1$ with $\Gamma_1 > 0$ in the low-density limit.
    The states corresponding to the indices (1)--(11) are summarized in Fig.~\ref{fig:states}.
  }
\end{figure}

\subsection{Quantum phase diagram at the field-induced QCP}
For $\mu < 0$ (gapped spectrum), $E_\text{eff}$ is simply minimized by $\rho^{}_{\pm\mathbf{Q}_n} = 0$ for $\forall n$; i.e., the solution  is a quantum paramagnet.
The instability of the Bose-Einstein condensation occurs at $\mu = 0$.
Figure~\ref{fig:phase-diagram1} shows a phase diagram obtained by minimizing $E_\text{eff}$ as a function of the effective interactions for a small constant value of the total density $\rho = \sum_{n}(\rho^{}_{\mathbf{Q}_n} + \rho^{}_{-\mathbf{Q}_n})$.
Here, we assume $\Gamma_1 > 0$, which is usually the case in antiferromagnets, where the boson-boson interaction is repulsive.
We also assume that $\lvert{\Gamma_5}\rvert$ is small relative to the others, which turns out to be the case in the microscopic calculation that we discuss later.

\begin{figure}[t]
  \includegraphics[width=0.9\hsize]{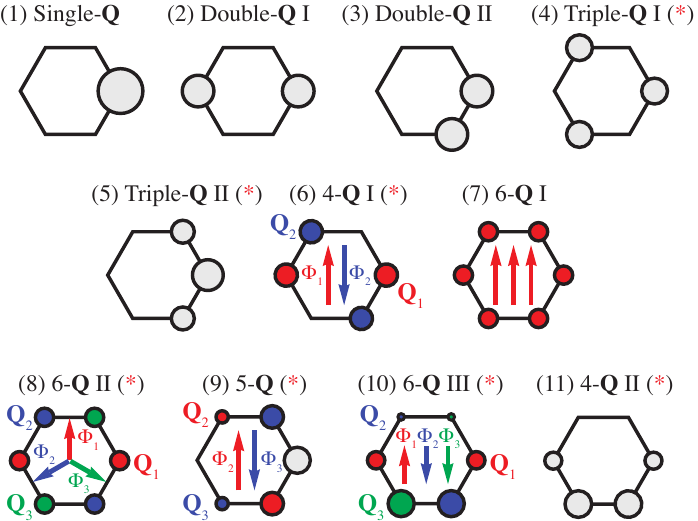}
  \caption{%
    \label{fig:states}%
    Schematic condensate distribution in momentum space for
    the single- and multi-$\mathbf{Q}$ states that appear at the BEC-QCP.
    The arrows representing $\Phi_n$ variables are shown only for states with some conditions on $\Phi_n$.
    The states marked with ($^\ast$) are vortex crystals (see text).
  }
\end{figure}

The phase diagram shows  a plethora of multi-$\mathbf{Q}$ orderings for $\Gamma_3 \simeq \Gamma_4 \lesssim \Gamma_1$ and $\Gamma_2 \lesssim \Gamma_1$. 
This condition implies that the single-$\mathbf{Q}$ state is not favored (because of the dominating $\Gamma_1$ vertex). It also implies that the effective interactions  between different modes, $\Gamma_{2}$, $\Gamma_{3}$, and $\Gamma_{4}$, are highly frustrated.
An additional condition $\Gamma_5 > 0$  leads to a bifurcation of exotic multi-$\mathbf{Q}$ states (No.~$6$ and Nos.~$8$--$11$).
Figure~\ref{fig:states} shows the corresponding schematic condensate distributions in momentum space. As we describe below, these states are magnetic vortex crystals, in which vortex cores of $xy$-spin components form a regular lattice (see below).
On the other hand, the condition $\Gamma_5 < 0$ almost exclusively favors the $6$-$\mathbf{Q}$~I BEC state where $\Phi_1 = \Phi_2 = \Phi_3$ (No.~$7$ in Fig.~\ref{fig:states}). This state corresponds to a coplanar state, which is not a vortex crystal. 
The reason for this contrast between $\Gamma \gtrless 0$ is an additional phase frustration that appears only when $\Gamma_5 > 0$ [see Eq.~\eqref{eq:GL}].

\begin{figure}[t]
  \includegraphics[width=0.98\hsize]{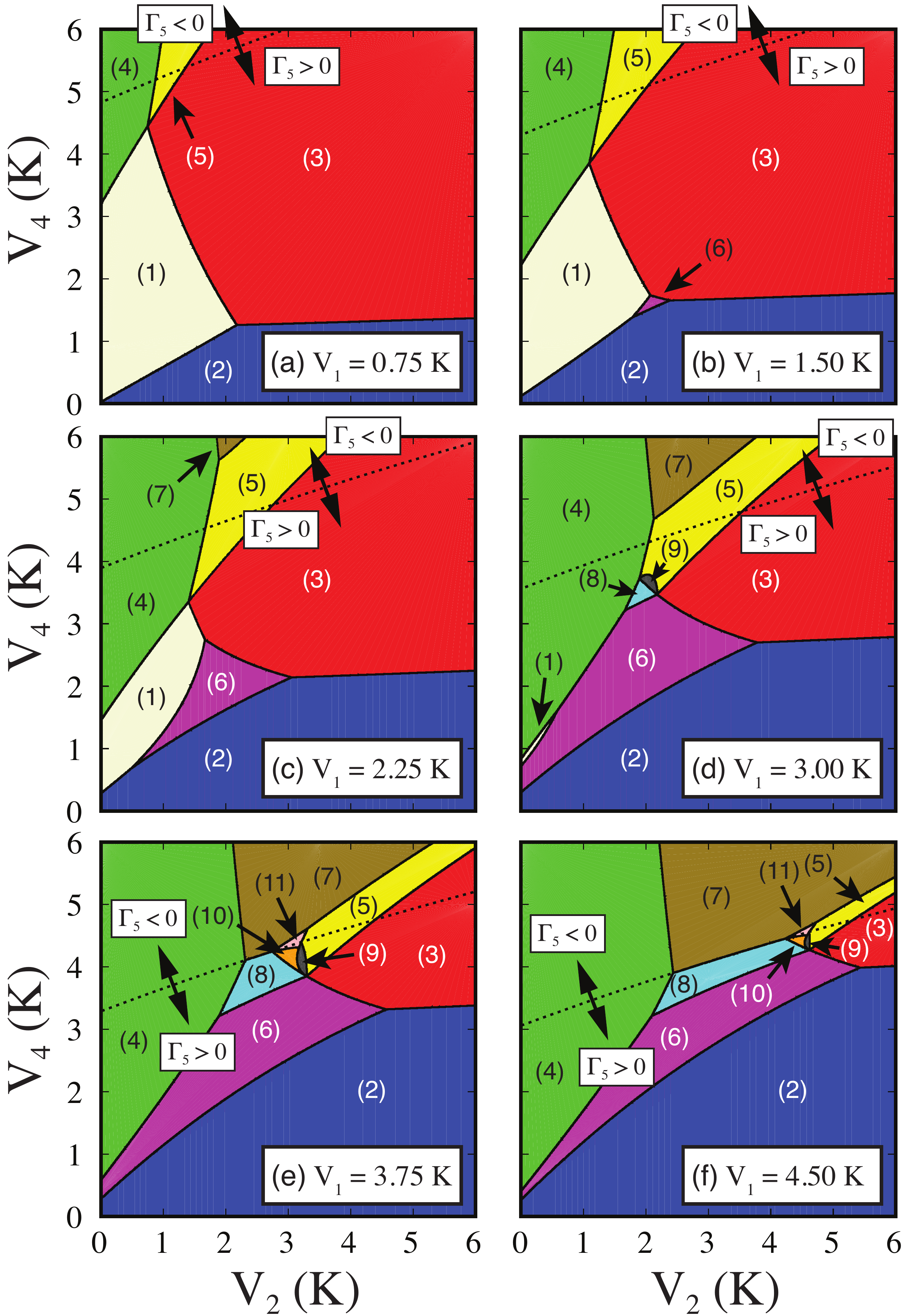}
  \caption{%
    \label{fig:phase-diagram2}%
    Phase diagram near the BEC-QCP for a system with the single-boson dispersion of Ba$_3$Mn$_2$O$_8$ as a function of $V_1$, $V_2$, and $V_4$. 
    The dashed line indicates a boundary across which the $\Gamma_5$ vertex changes sign.
    The states corresponding to the indices (1)--(11) are summarized in Fig.~\ref{fig:states}.
  }
\end{figure}

The phase diagram becomes rather simple away from the region that we described above.
We can easily narrow down an optimal state in the  limit of strong density-density interactions. 
For instance, the optimal state for $\Gamma_3 \gg \Gamma_1$ belongs to the set $\{\text{No.~}1,\text{No.~}2,\text{No.~}3\}$, which avoids this repulsive coupling.
Similarly, the optimal state for $\Gamma_2 \gg \Gamma_1$ and $\Gamma_4 \gg \Gamma_1$ belongs to $\{\text{No.~}1,\text{No.~}3,\text{No.~}4,\text{No.~}5\}$ and $\{\text{No.~}1,\text{No.~}2,\text{No.~}4\}$, respectively. In these limits, the small $\Gamma_5$ term is ineffective because the density prefactor becomes $0$ for all of the above candidate states.

Given this rich phase diagram, it is important to determine if these multi-$\mathbf{Q}$ BEC states can be realized under  realistic conditions. 
As we already mentioned, for a field-driven QCP of the Bose-Einstein condensation, we can compute the GL coefficients from a microscopic model under control; i.e., $E_\text{eff}$~\eqref{eq:GL} is not a phenomenological mean-field theory. 
For definiteness, we consider the spin Hamiltonian that is proposed for Ba$_3$Mn$_2$O$_8$.\cite{Uchida2002High-field,Stone2008Singlet-Triplet,Samulon2008Ordered,Samulon2009Asymmetric,Samulon2010Anisotropic,Suh2011Nonuniversal} 
We assume $\mathbf{H} \parallel c$ so that the U(1) symmetry is almost exact.
In the following, we fix $\omega_\mathbf{k}$ to the dispersion obtained from inelastic neutron scattering experiments,\cite{Stone2008Singlet-Triplet} while $V_1$, $V_2$, and $V_4$, which parametrize $V_\mathbf{q}$, are regarded as free parameters because they are not constrained by neutron scattering experiments.\footnote{While the inelastic neutron scattering experiments can provide information on the hopping parameters, the off-site interaction potentials are normally more difficult to measure; the high-$T$ susceptibility measurement provides a clue, which, however, does not fully constrain the model.}
We focus on the region $0 \le V_1,\,V_2,\,V_4 \ll J_0 \approx 19\mathrm{K}$ corresponding to the weakly coupled spin dimers with repulsive triplon-triplon interactions (appropriate for antiferromagnetic compounds like Ba$_3$Mn$_2$O$_8$).

We produce the phase diagram shown in Fig.~\ref{fig:phase-diagram2} by computing $\Gamma_{1 \le \nu \le 5}$ as a function of these microscopic interaction parameters. The density-density interaction vertices, $\Gamma_{1 \le \nu \le 4}$, are positive in the investigated region, as expected from the repulsive bare interactions.
The $\Gamma_5$ vertex has a much smaller amplitude.
Most importantly, we confirm the plethora of multi-$\mathbf{Q}$ phases in this microscopic model. They are stabilized when $V_1 \gtrsim 1.5\mathrm{K}$ and some other conditions on $V_2 \simeq V_3$ are fulfilled.
We also observe that the sign change of the $\Gamma_5$ vertex can induce an instability towards the coplanar $6$-$\mathbf{Q}$~I state (No.~7), as indicated by the dashed line in Figs.~\ref{fig:phase-diagram2}(e) and \ref{fig:phase-diagram2}(f).

\section{Vortex crystals}

\subsection{Spin configurations}
The BEC state for $\mu > 0$ can be approximated by
\begin{align}
  \left\langle{b_i}\right\rangle \sim \sum_{n=1}^{3}
  \left(
  \rho^{}_{\mathbf{Q}_n} e^{i\phi^{}_{\mathbf{Q}_n}} e^{i\mathbf{Q}_n \cdot \mathbf{r}_i}
  + \rho^{}_{-\mathbf{Q}_n} e^{i\phi^{}_{-\mathbf{Q}_n}} e^{-i\mathbf{Q}_n \cdot \mathbf{r}_i}
  \right).
\end{align}
The actual spin configuration for either a dimerized compound or a generic quantum antiferromagnet near the saturation field follows from the spin-boson transformations~\cite{Matsubara1956Lattice} (see Appendix~\ref{app:spinconfig}). 
The simplest state is the well-known single-$\mathbf{Q}$ state (No.~$1$), which is an $xy$ spiral with a uniform magnetization along the field direction. 
The double-$\mathbf{Q}$~I state (No.~$2$) has also been well studied,\cite{Nikuni1995Hexagonal,Ueda2009Magnon} and it is a coplanar state with a one-dimensional modulation (``fan''). 
In general, the other multi-$\mathbf{Q}$ spin states correspond to richer spin structures.
In particular, states Nos.~$4$--$6$, and Nos.~$8$--$11$ are vortex crystals, whose emergent lattice parameter is controlled by $\lvert{\mathbf{Q}_n}\rvert^{-1}$ when $\lvert{\mathbf{Q}_n}\rvert \ll 1$.

\begin{figure}[t]
  \includegraphics[width=\hsize]{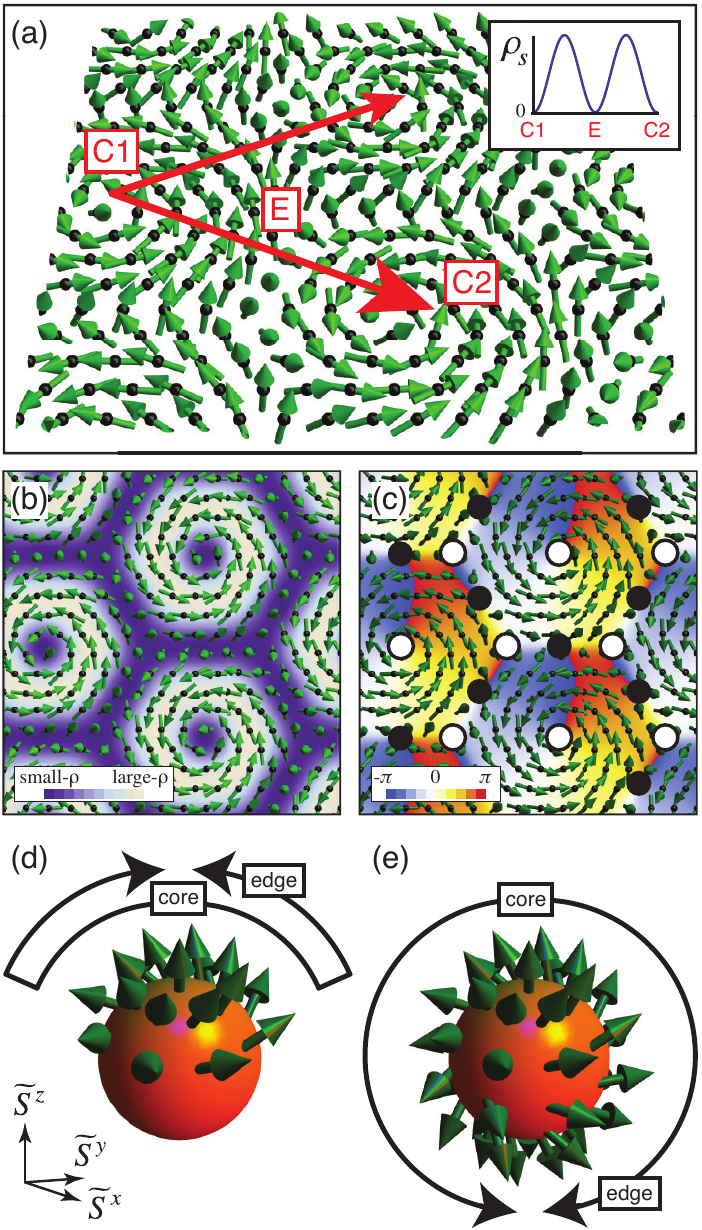}
  \caption{%
    \label{fig:6-QIIPi}%
    (a) The $6$-$\mathbf{Q}$~II BEC state for $\kappa = +1$, $\Theta = \pi$, and $\lvert{\mathbf{Q}_n}\rvert \ll 1$. 
    The inset shows a modulation of the pseudo-spin $z$-component along C1-E-C2 that corresponds to a vortex lattice unit spacing.
    (b) Contour plot showing the boson density distribution.
    (c) Contour plot showing the boson phase distribution. The open (filled) circles indicate vortices (anti-vortices).
    (d) Spin structure of a single vortex mapped on a sphere in the spin space, which is compared to (e) the same plot of a skyrmion.
  }
\end{figure}
\begin{figure}[t]
  \includegraphics[width=\hsize]{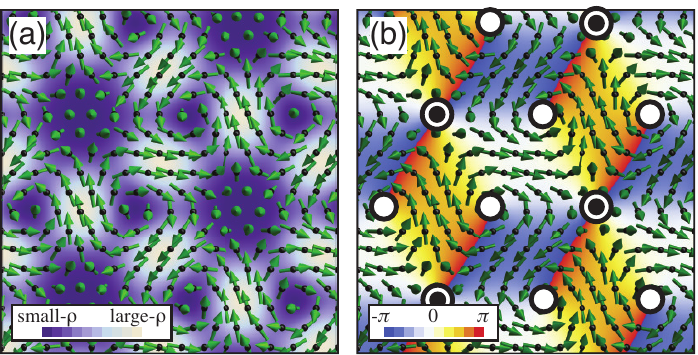}
  \caption{%
    \label{fig:6-QIIZR}%
    The $6$-$\mathbf{Q}$~II BEC state for $\kappa = +1$, $\Theta = 0$, and $\lvert{\mathbf{Q}_n}\rvert \ll 1$ showing  
    (a) the boson density and (b) boson phase distributions.
    The open (filled double) circles indicate vortices (double-antivortices).
  }
\end{figure}
\begin{figure}
  \includegraphics[width=\hsize]{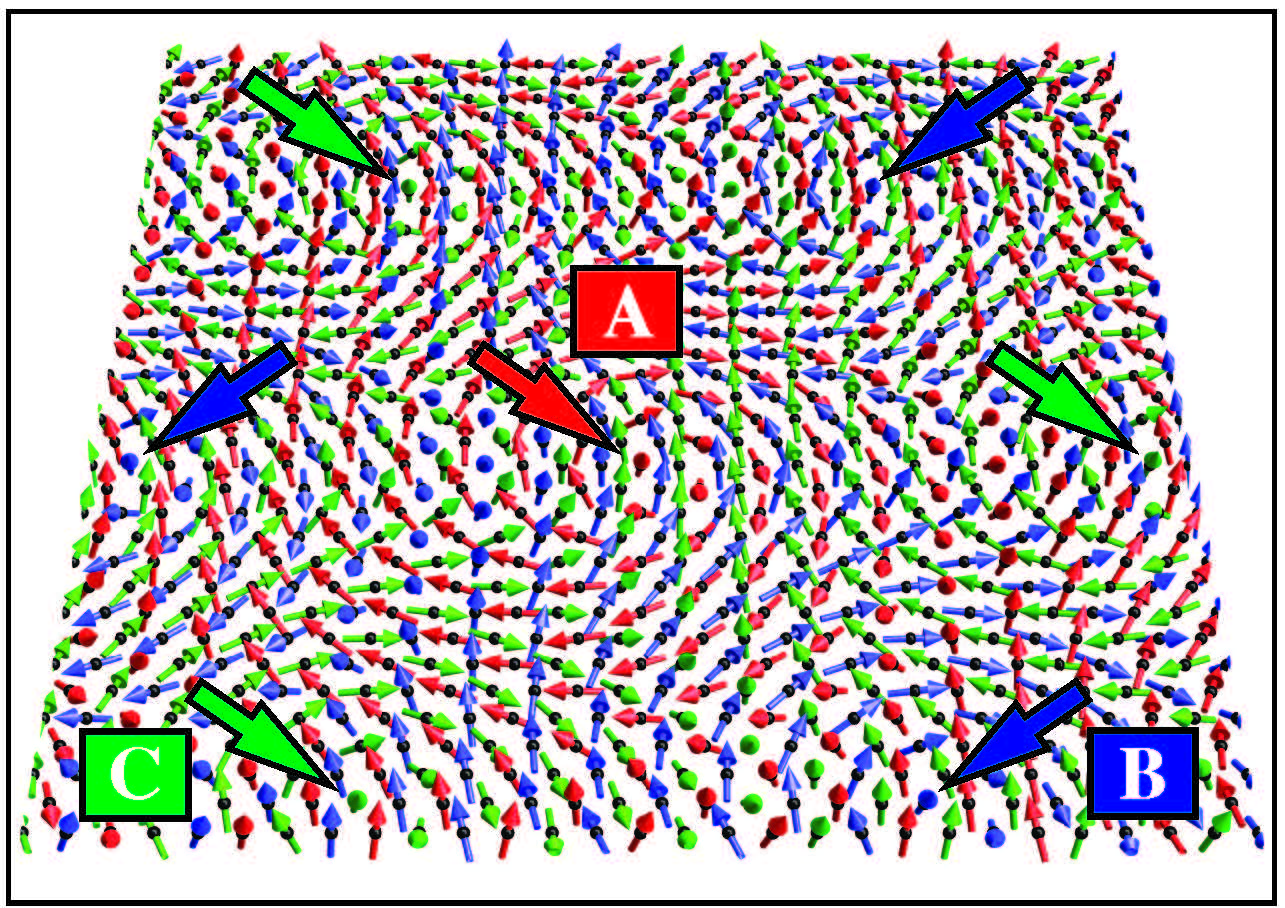} 
  \caption{%
    \label{fig:6-QIIZR-BMO-Q}%
    Three-sublattice structure of the $6$-$\mathbf{Q}$~II state, where $\Theta = \pi$, $\kappa = +1$, and $\pm\mathbf{Q}_{1 \le n \le 3}$ coincide with the triplon dispersion minima of Ba$_3$Mn$_2$O$_8$ (see Fig.~\ref{fig:dispersion}).
    The arrows indicate locations of the vortex cores of each sublattice $A$, $B$, and $C$.
  }
\end{figure}
\begin{figure*}
  \includegraphics[width=\hsize]{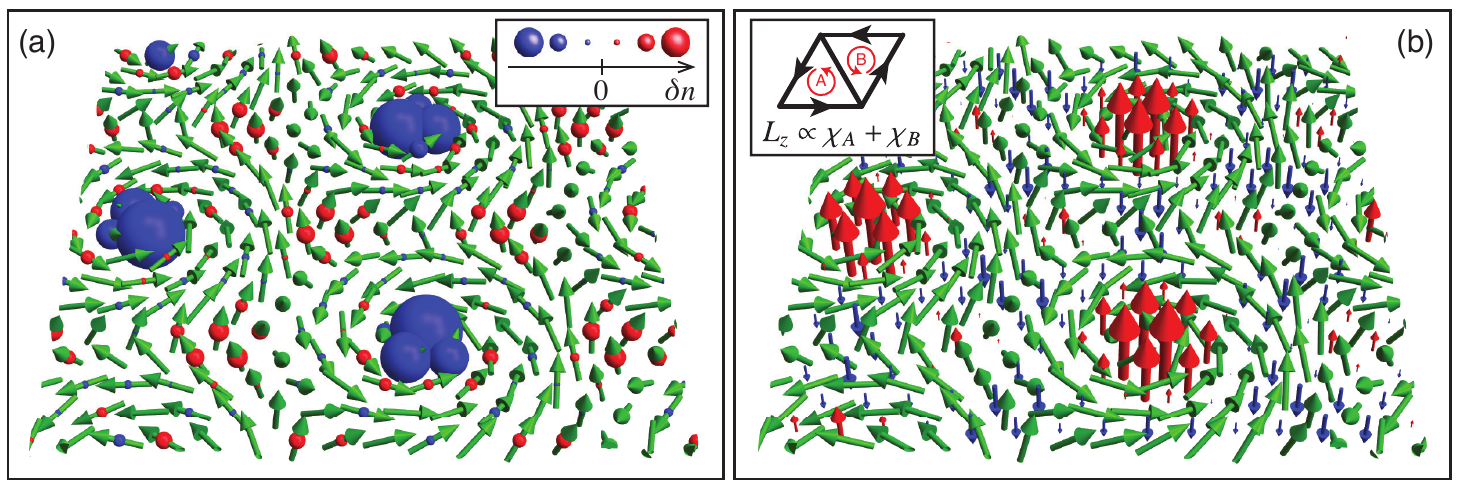}
  \caption{%
    \label{fig:dneff}
    (a) Charge-density wave induced by the $6$-$\mathbf{Q}$~II state with $\kappa = +1$, $\Theta = 0$, and $\lvert{\mathbf{Q}_n}\rvert \ll 1$. 
    (b) Distribution of orbital magnetic moments induced by the same phase.
    Orbital currents nearly cancel each other on the common bonds of two neighboring plaquettes (see the inset).
  }
\end{figure*}

To illustrate the main characteristics of vortex crystals, we take the $6$-$\mathbf{Q}$~II BEC state (No.~$8$) and describe its spin structure in some detail.
The condensates of this state occupy all of the six degenerate single-particle states, $\mathbf{k} = \pm \mathbf{Q}_{1\le n\le3}$, with equal amplitudes.
The relative phases $\Phi_{n+1} - \Phi_{n}$ ($\Phi_4 \equiv \Phi_1$) take the values  $\pm 2\pi/3$ because of the positive $\Gamma_5$ term in Eq.~\eqref{eq:GL}:
\begin{align}
  \Phi_n =  \frac{2n\kappa\pi}{3} + \text{const.},~\kappa = \pm 1.
  \label{eq:Phi:6-Q}
\end{align}
The spin structure is derived from
\begin{align}
  \left\langle{b_i}\right\rangle \sim \sqrt{\frac{2\rho^{}}{3}}
  \sum_{n=1}^{3}
  \cos\left[
    \mathbf{Q}_n \cdot \left(\mathbf{r}_i - \mathbf{r}^\ast\right) + \delta_{n,3} \frac{\Theta}{2}
    \right]
  e^{i\Phi_n/2},
  \label{eq:6-Q}
\end{align}
which is exact in the low-density limit.
Here, we have chosen a shift $\mathbf{r}^\ast$ to emphasize the other phase parameter,
\begin{align}
  \Theta = \sum_{n=1}^{3} \left(\phi^{}_{\mathbf{Q}_n} - \phi^{}_{-\mathbf{Q}_n}\right),
\end{align}
which is invariant under the U(1) group of global spin rotations along the field axis.
$\Theta$ can be equal to $0$ or $\pi$ depending on the sign of the  three-body scattering vertex $\Gamma_6$ which leads to the following contribution to the Hamiltonian (in which, however, we do not explicitly evaluate $\Gamma_6$):
\begin{align}
  \mathpzc{H}_{\,6} 
  &\sim \Gamma_6 
  \left(
  b^\dagger_{\mathbf{Q}_1} b^\dagger_{\mathbf{Q}_2} b^\dagger_{\mathbf{Q}_3}  
  b^{}_{-\mathbf{Q}_1} b^{}_{-\mathbf{Q}_2} b^{}_{-\mathbf{Q}_3}  
  + \text{H.c.}
  \right)
  \notag\\
  &\sim 
  2\Gamma_6
  \sqrt{\prod_n\rho^{}_{\mathbf{Q}_n}\rho^{}_{-\mathbf{Q}_n}}
  \cos\Theta.
  \label{eq:three-body}
\end{align}
Note that this contribution conserves momentum  because $\mathbf{Q}_1 + \mathbf{Q}_2 + \mathbf{Q}_3 = 0$.

In Fig.~\ref{fig:6-QIIPi}, we show the spin configuration of this BEC state for $\kappa = +1$, $\Theta = \pi$, and $\lvert{\mathbf{Q}_{n}}\rvert \ll 1$ on a given triangular lattice layer. 
The spin configuration corresponds to a crystal of magnetic vortices, and this is essentially invariant along the $c$ axis.
Here, we adopt a pseudospin representation $\widetilde{\mathbf{S}}$, i.e., $\lvert{\widetilde{\uparrow}}\rangle_i \Leftrightarrow \lvert{\varnothing}\rangle_i$ (i.e., a singlet dimer) and $\lvert{\widetilde{\downarrow}}\rangle_i \Leftrightarrow b^{\dagger}_i \lvert{\varnothing}\rangle$ (a triplet dimer), where the amount of canting relative to the $z$ axis corresponds to the local condensate density and the azimuth angle is equal to the local boson phase (see Appendix~\ref{app:spinconfig}).
Figure~\ref{fig:6-QIIPi}(b) shows that the condensate density is suppressed near each vortex core and becomes equal to $0$ right at the center of the core.
The density gradually increases away from a core, and the $xy$ spin components wind around the core. The phase change in winding around the core is $2\kappa\pi$, depending on the two branches, $\kappa = \pm 1$, of Eq.~\eqref{eq:Phi:6-Q}. Thus, the vortex crystal is a chiral spin texture. 
The condensate density starts decreasing again beyond a certain radius from the core because 
the ``edges'' that separate different  vortices are low-density regions which form a honeycomb lattice.
These crystallized vortices are not topological defects but thermodynamic stable states similar to the skyrmion crystals observed in B20 compounds~\cite{Muhlbauer2009Skyrmion,Yu2010Real-space,Seki2012Observation} and in triangular lattice models for classical spins~\cite{Okubo2012Multiple} or to the crystals of magnetic $Z_2$ vortices~\cite{Rousochatzakis:arXiv1209.5895} and solitons~\cite{Kamiya2012Formation,Selke1988ANNNI,Bak1982Commensurate} obtained in different contexts.The Abrikosov lattice in the type-II superconductors is another example of this kind.~\cite{Abrikosov1956Magnetic}

Another important observation is that the net spin-solid-angle wrapped by a single vortex is always $0$ [see Fig.~\ref{fig:6-QIIPi}(d)].
Starting from a vortex core, $\langle{\widetilde{\mathbf{S}^{}_{}}}\rangle$ wraps some fraction of a sphere from the ``north pole.'' However, $\langle{\widetilde{S}^z}\rangle$ starts to 
increase again beyond a threshold radius, meaning that $\langle{\widetilde{\mathbf{S}^{}_{}}}\rangle$ starts unwrapping the sphere. After including the whole contribution up to the vortex edge, the contribution from inside the threshold radius is exactly cancelled.
Therefore, these structures are not skyrmions, $\pi_2(S^2)$, which wrap the full solid angle of a sphere [Fig.~\ref{fig:6-QIIPi}(e)].
However, they are certainly Abelian vortices $\pi_1(S^1)$ because of the structure of the $xy$-spin components.
Similarly to the case of skyrmion crystals,\cite{Muhlbauer2009Skyrmion,Yu2010Real-space,Okubo2012Multiple,Seki2012Observation} 
the Abelian vortex crystals are also characterized by regularly spaced vortex cores.
To elucidate this property, we show a contour plot of the boson phase, $\arg\langle{b_i}\rangle$, for the same spin configuration in Fig.~\ref{fig:6-QIIPi}(c). The endpoints of branch cuts between $\arg\langle{b_i}\rangle=\pm\pi$ indicate the locations of vortex cores, which indeed form a lattice structure. 
Another interesting observation is the presence of the vortex cores on the edges, which are less evident in the other plots.
The net vorticity is zero because of these additional vortices; i.e., there is no branch cut that is connected to the infinity point.

Figure~\ref{fig:6-QIIZR} shows the other variant, $\Theta = 0$, of the $6$-$\mathbf{Q}$~II state ($\kappa = +1$). The high-density regions form a kagome lattice.
The vortex cores are located at the center of the faces of this kagome lattice. 
Single vortices (winding number equal to $\kappa$) are located at the center of the smaller faces, while double-antivortices (winding number $-2\kappa$) are located at the center of the bigger faces.
The phase contour plot shown in Fig.~\ref{fig:6-QIIZR}(b) elucidates two branch cuts coming out of one double-antivortex, both of which are connected to a unitary vortex. This shows that there is no net vorticity for $\Theta = 0$.

So far we have described the case  $\lvert{\mathbf{Q}_n}\rvert \ll 1$. 
Now, we will briefly comment  on the case relevant for  Ba$_3$Mn$_2$O$_8$, where the $\mathbf{Q}$ vectors are located very close to the K point of the 2D Brillouin zone (see Fig.~\ref{fig:dispersion}).
The proximity to the K point  leads to a local spin structure that resembles a three-sublattice order. 
The small deviation $\Delta {\mathbf{Q}_n}$ induces a long-wavelength spin modulation, which results in the vortex crystal on each of the three sublattices (Fig.~\ref{fig:6-QIIZR-BMO-Q}). The superlattice spacing $\lambda$  is proportional to the inverse of this deviation, $\lambda \propto 1/|\Delta {\mathbf{Q}_n}|$.

\subsection{Dielectric properties and orbital currents}

The complex noncoplanar spin structure of the vortex crystals can lead to nontrivial dielectric properties.
Magnetoelectric behavior appears naturally in these structures because they locally break most of the symmetries of the underlying crystal.
For example, the local vector chirality of the vortex crystal will naturally affect the locations of the nonmagnetic ions that mediate superexchange through the so-called inverse Dzyaloshinskii-Moriya mechanism.\cite{Cheong2007Multiferroics}
Even without considering the spin-lattice coupling, spin textures that break the equivalence between bonds (bond ordering) or develop a finite  scalar spin chirality  induce purely electronic charge effects resulting from virtual processes on frustrated plaquettes.\cite{Bulaevskii2008electronic,Kamiya2012Multiferroic} 
Below, we demonstrate that this is indeed the case for our vortex crystals.

For definiteness, we consider a spin-$1/2$ quantum antiferromagnet very close to the saturation field $H = H_\text{sat}$ on the nondimerized variant of the period-$3$ stacked triangular lattice, as shown in the inset of Fig.~\ref{fig:dispersion}. 
Our theory also applies to this situation if we just identify the pseudospin $\widetilde{\mathbf{S}}$ with a real spin-$1/2$ operator. In that case, $\lvert{\varnothing}\rangle$ becomes the fully polarized spin state and $b^\dagger_\mathbf{r}$ is an operator that flips a spin on site $\mathbf{r}$. 
By neglecting contributions from the interlayer electron hopping for the sake of simplicity, the effective electronic-charge-density operator is $1+\delta {\tilde n}_\mathbf{r}$, with~\cite{Bulaevskii2008electronic} 
\begin{align}
  \delta {\tilde n}_\mathbf{r} \propto 
  \sum_{0\le\eta<6} \left(
  \mathbf{S}_{\mathbf{r}} - \mathbf{S}_{\mathbf{r}+\Delta\mathbf{r}^{}_{\eta+1}}
  \right)
  \cdot \mathbf{S}_{\mathbf{r}+\Delta\mathbf{r}^{}_\eta},
  \label{eq:dneff}
\end{align}
where $\Delta\mathbf{r}^{}_{0 \le \eta < 6}$ runs counterclockwise over the displacement vectors to the NN sites of the triangular lattice layer.
The lowest-order contribution to $\delta {\tilde n}_\mathbf{r}$ is of third order in the electron hopping. The three hopping processes must close a triangular plaquette. Equation~\eqref{eq:dneff} is obtained by adding  contributions from the six triangles connected to the site $\mathbf{r}$.
Figure~\ref{fig:dneff}(a) shows the distribution of $\delta {\tilde n}_\mathbf{r}$ in the case of the $6$-$\mathbf{Q}$~II vortex crystal (No.~8) with $\Theta = \pi$. The results are insensitive to $\kappa = \pm 1$.
We find a charge-density wave resulting from the long-wavelength modulation of the spin texture. 
This charge modulation should be, in principle, detectable by x-ray measurements.

As it was also pointed out in Ref.~\onlinecite{Bulaevskii2008electronic},  the scalar spin chirality on each triangle is a manifestation of a local electronic current (orbital current) that circulates around the triangular plaquette.
The effective orbital current operator for a triangle $\mathbf{r}_1$-$\mathbf{r}_2$-$\mathbf{r}_3$ is~\cite{Bulaevskii2008electronic}
\begin{align}
 {\tilde I} (\mathbf{r}_1, \mathbf{r}_2, \mathbf{r}_3) \propto \chi^{}_{1,2,3} = \left(\mathbf{S}_{\mathbf{r}_1} \times \mathbf{S}_{\mathbf{r}_2}\right) \cdot \mathbf{S}_{\mathbf{r}_3}.
\end{align}
The orbital current produces an orbital magnetic moment $\mathbf{L}$ normal to the plaquette.
Once again, by only considering the effect of the in-plane electronic hopping in the $6$-$\mathbf{Q}$~II vortex crystal with $\Theta = \pi$ and $\lvert{\mathbf{Q}}\rvert \ll 1$, we obtain that contributions of neighboring plaquettes tend to cancel each other,  but the vortex crystal still sustains  orbital currents well beyond the scale of the lattice spacing [see Fig.~\ref{fig:dneff}(b)]. 

\section{Anisotropy effects}

\subsection{Symmetric exchange anisotropy}

\begin{figure}[t]
  \includegraphics[width=\hsize]{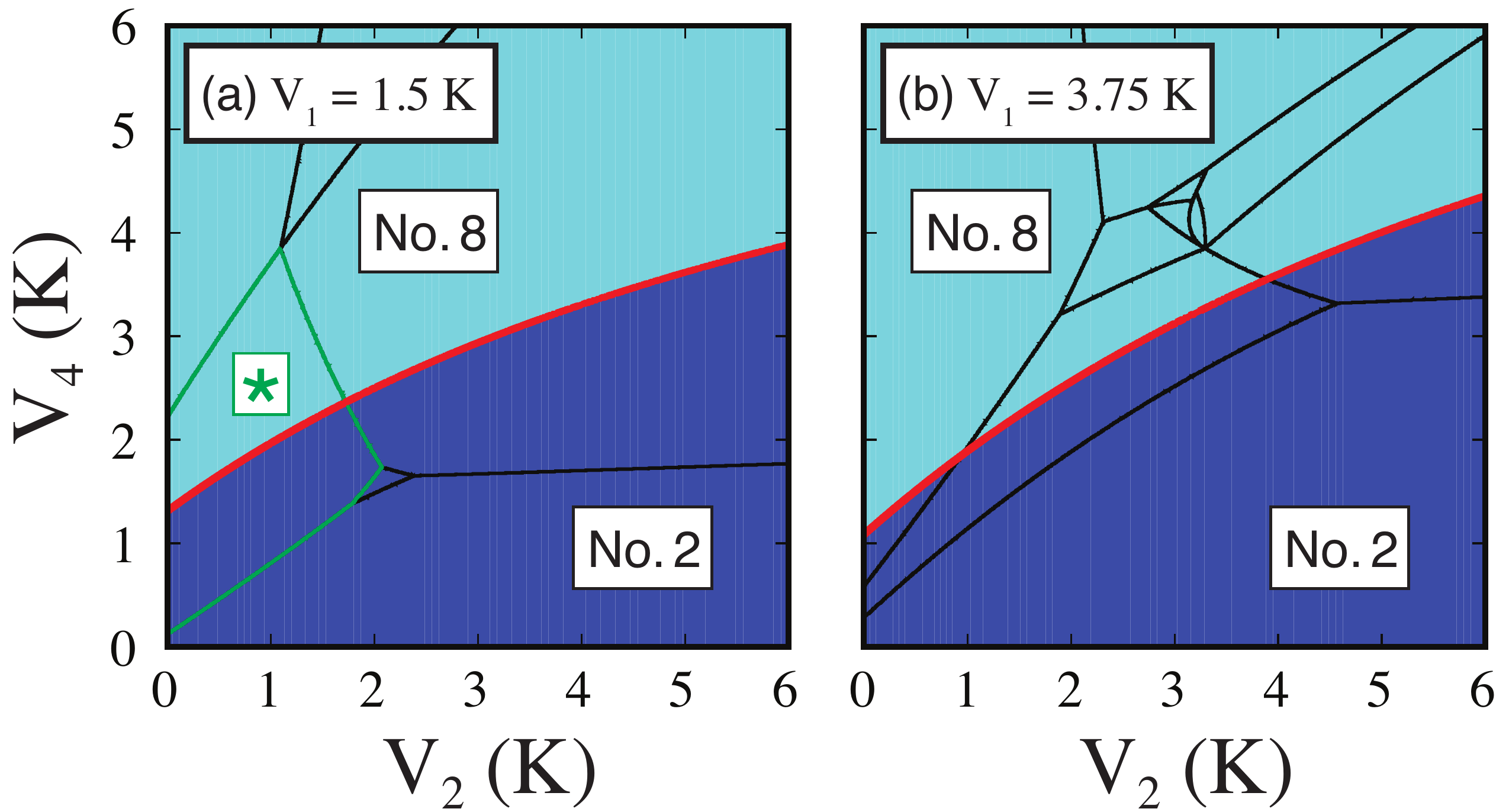}
  \caption{%
    \label{fig:phase-diagram3}%
    Phase diagram at the field-induced QCP for the same system as in Fig.~\ref{fig:phase-diagram2} after including the symmetric exchange anisotropy $\mathpzc{H}_{\,A,6}$ in Eq.~\eqref{eq:aniso1}.
    The phase boundary corresponds to the condition $2(\Gamma_1 + \Gamma_2 - \Gamma_3 - \Gamma_4) + \Gamma_5 = 0$.
    For comparison, phase boundaries only due to the  interaction terms are also shown; 
    the region indicated by ($\ast$) is the single-$\mathbf{Q}$ phase.
  }
\end{figure}

The U($1$) symmetry of the spin Hamiltonian [Eq.~\eqref{eq:ham}] results from an approximation. 
In real magnets there are several anisotropy terms that break any continuous symmetry. 
Contributions from exchange anisotropy are proportional to  $\rho$. Thus,  no matter how small their amplitude might be, they  dominate the interaction effects ($\propto \rho^2$) in the low-density limit, i.e., close enough to the field-induced QCP.
We consider the in-plane  symmetric exchange anisotropy, $\mathpzc{H}_{\,A,6} = (J_{A,6}/2) \sum_{\mathbf{r},\eta} (e^{-2\eta\pi i/3} b^{\dagger}_{\mathbf{r}} b^{\dagger}_{\mathbf{r}+\Delta\mathbf{r}_\eta} + \text{H.c.})$, which can be induced either by the relativistic spin-orbit coupling or by dipole-dipole interactions. This contribution breaks the U($1$) symmetry of $\mathpzc{H}$ down to the six-fold symmetry of the triangular lattice.

After taking the long wavelength limit of $\mathpzc{H}_{\,A,6} $ we obtain
\begin{align}
  \mathpzc{H}_{\,A,6} 
  &\sim \frac{\gamma J_{A,6}}{N} \sum_{n=1}^{3}
  \left(
  e^{2(n-1)\pi i/3} b^{\dagger}_{\mathbf{Q}_n} b^{\dagger}_{-\mathbf{Q}_n} + \text{H.c.}
  \right)
  \notag\\
  &\sim 2 \gamma J_{A,6} \sum_{n=1}^{3}
  \sqrt{\rho^{}_{\mathbf{Q}_n} \rho^{}_{-\mathbf{Q}_n}} \cos \left( \Phi_n - \frac{2(n-1)\pi}{3} \right),
  \label{eq:aniso1}
\end{align}
where $\gamma$ is a constant prefactor.
Equation~\eqref{eq:aniso1} suggests that the phases $\Phi_{1 \le n \le 3}$ must be adjusted to $\Phi_{n} = 2(n-1)\pi/3$ [$\Phi_{n} = (2n+1)\pi/3$] for $\gamma J_{A,6} < 0$ ($\gamma J_{A,6} > 0$) in the low-density limit.
In addition, the pairs $\mathbf{k} = \pm \mathbf{Q}_n$ ($1 \le n \le 3$) must have the same condensate amplitudes if they are finite.
At this level, there are three degenerate solutions: the double-$\mathbf{Q}$ fan state (No.~2), the $4$-$\mathbf{Q}$ state similar to No.~$6$ but with the above condition for $\Phi_{n}$, and the $6$-$\mathbf{Q}$~II state (No.~8). This degeneracy is lifted by the boson-boson interaction [Eq.~\eqref{eq:GL}].
The final result is that the $6$- (double-)$\mathbf{Q}$ state has the lowest energy if $2(\Gamma_1 + \Gamma_2 - \Gamma_3 - \Gamma_4) + \Gamma_5 > 0$ ($< 0$) and there is no chance for the $4$-$\mathbf{Q}$ state to be a stable solution.
Figure~\ref{fig:phase-diagram3} shows that $\mathpzc{H}_{\,A,6}$ significantly enlarges the $6$-$\mathbf{Q}$~II phase near the field-induced QCP.

Thermodynamic measurements on Ba$_3$Mn$_2$O$_8$ show two different phases in the vicinity of $H = H_{c1}$ even for $\mathbf{H} \parallel c$.\cite{Samulon2010Anisotropic} A narrow ``phase II'' appears right above $H_{c1}$, while a broad  single-$\mathbf{Q}$ spiral ``phase I''  extends over a much bigger window of magnetic fields inside the dome of  ordered phases.
Because $\mathpzc{H}_{\,A,6}$ becomes  ineffective sufficiently away from the QCP, 
the broad window of phase I suggests that the interaction parameters $V_1$, $V_2$, and $V_4$ favor the single-$\mathbf{Q}$ state. 
As is shown in Fig.~\ref{fig:phase-diagram3}, the single-$\mathbf{Q}$ phase obtained without taking $\mathpzc{H}_{\,A,6}$ into account is typically crossed by the line of the condition $2(\Gamma_1 + \Gamma_2 - \Gamma_3 - \Gamma_4) + \Gamma_5 = 0$.
This observation indicates that when the angle between $\mathbf{H}$ and the $c$ axis is small, either the double-$\mathbf{Q}$ fan state (No.~2) or the type-II vortex crystal No.~8 stabilized by $\mathpzc{H}_{\,A,6}$ could explain phase II. This prediction can be verified by performing nuclear magnetic resonance or neutron scattering experiments right above $H = H_{c1}$ for $\mathbf{H} \parallel c$. 

\subsection{Uniaxial anisotropy and the magnetic field tilted from the symmetry axis}

\begin{figure}
  \includegraphics[width=\hsize]{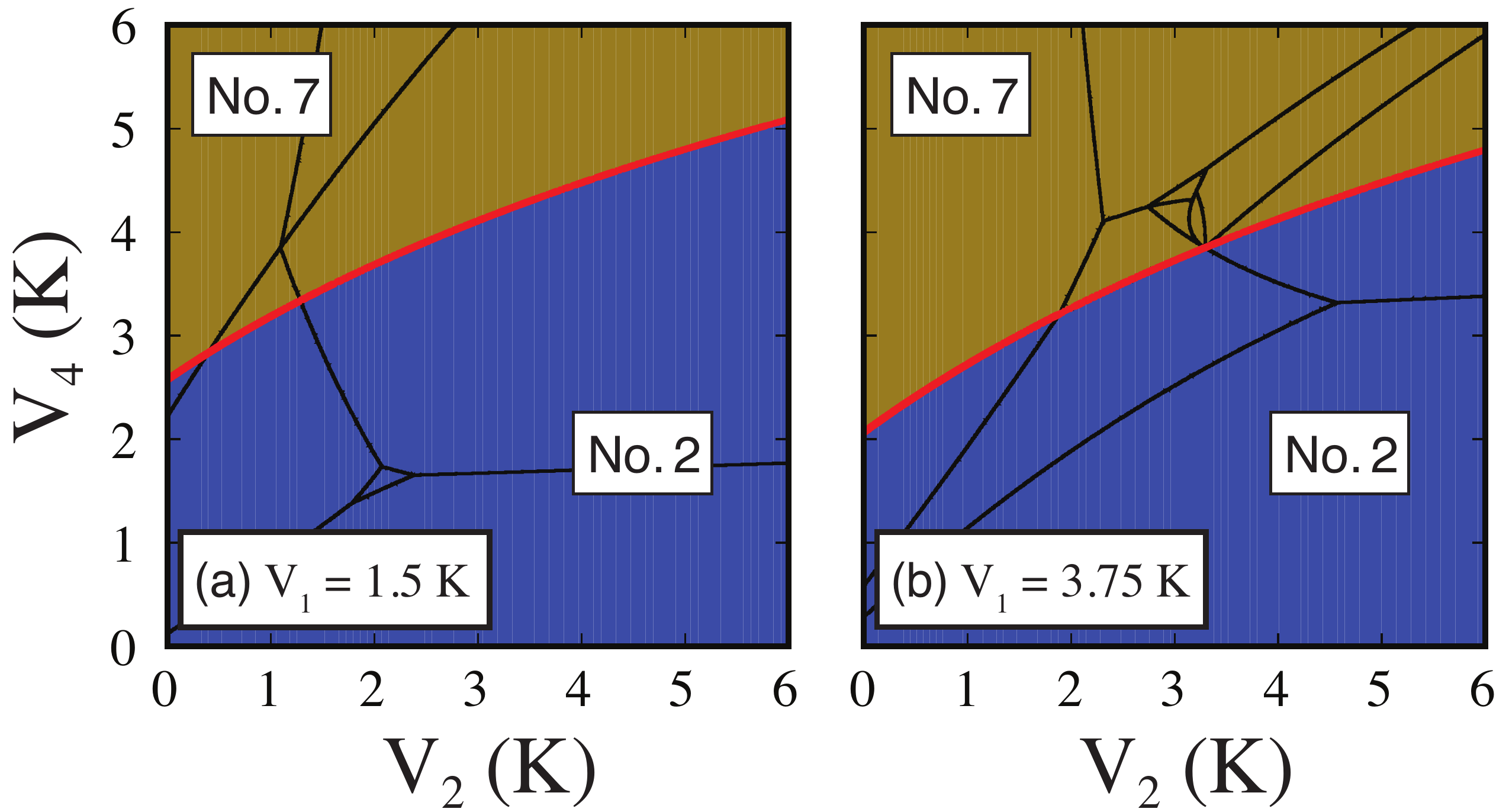}
  \caption{%
    \label{fig:phase-diagram4}%
    Phase diagram at the field-induced QCP for the same system as in Fig.~\ref{fig:phase-diagram2} after including the effect of exchange anisotropy $\mathpzc{H}_{\,A,2}$ in Eq.~\eqref{eq:aniso2}.
    The phase boundary corresponds to the condition $\Gamma_1 + \Gamma_2 - \Gamma_3 - \Gamma_4 - \Gamma_5 = 0$.
    For comparison, phase boundaries only due to the  interaction terms are also shown.
  }
\end{figure}

If the system has an uniaxial magnetic anisotropy along the $c$ axis, the U($1$) symmetry requires a fine-tuning of the magnetic-field direction along this symmetry axis. In other words, a different symmetry-breaking term appears when $\mathbf{H}$ is tilted from the $c$ axis.
Recently, field-angle-dependent phases and phase transitions have become an interesting subject in frustrated magnets such as Ba$_3$Mn$_2$O$_8$~\cite{Samulon2008Ordered,Samulon2010Anisotropic} and Ba$_3$CoSb$_2$O$_9$.\cite{Susuki2013Magnetization,Zhou2012Successive,Koutroulakis:arXiv1308.6331}

Below, we address the effects due to the uniaxial anisotropy and $\mathbf{H}$ away from the $c$ axis. We neglect the effect of $\mathpzc{H}_{\,A,6}$ in Eq.~\eqref{eq:aniso1} for the sake of simplicity, and maintain the quantization axis always along the field direction.
The additional term that appears in the long-wavelength limit is
\begin{align}
  \mathpzc{H}_{\,A,2} 
  &\sim \frac{\gamma' J_{A,2}}{N} \sum_{n=1}^{3}
  \left(
  b^{\dagger}_{\mathbf{Q}_n} b^{\dagger}_{-\mathbf{Q}_n} + \text{H.c.}
  \right)
  \notag\\
  &\sim 2 \gamma' J_{A,2} \sum_{n=1}^{3}
  \sqrt{\rho^{}_{\mathbf{Q}_n} \rho^{}_{-\mathbf{Q}_n}} \cos \Phi_n,
  \label{eq:aniso2}
\end{align}
where $\gamma'$ is a numerical prefactor.
This term breaks the U($1$) symmetry down to $Z_2$ symmetry. $\mathpzc{H}_{\,A,2}$ also scales as $\propto \rho$ and thus dominates the interaction terms in the low-density limit. Consequently, $\Phi_{1 \le n \le 3}$ are enforced to take the same value $\Phi = 0$, $\pi$ depending on the sign of its prefactor, and the pair $\mathbf{k} = \pm \mathbf{Q}_n$ ($1 \le n \le 3$) must have the same condensate amplitudes if they are finite.
As a result, $\mathpzc{H}_{\,A,2}$ leaves three degenerate states: the double-$\mathbf{Q}$ fan state (No.~2), a $4$-$\mathbf{Q}$ state similar to No.~6 but with $\Phi_{n} = \Phi$ ($= 0$, $\pi$), and the coplanar $6$-$\mathbf{Q}$~I state (No.~7).
The boson-boson interaction lifts this degeneracy: the $6$- (double-)$\mathbf{Q}$ state is stabilized for $\Gamma_1 + \Gamma_2 - \Gamma_3 - \Gamma_4 - \Gamma_5 > 0$ ($< 0$); see Fig.~\ref{fig:phase-diagram4}.
The condensate of the coplanar $6$-$\mathbf{Q}$~I state is described by Eq.~\eqref{eq:6-Q} with $\Phi_n = \Phi$.
Slightly different spin configurations are obtained depending on $\Theta = 0$, $\pi$, which is determined by the sign of the three-body scattering term in Eq.~\eqref{eq:three-body}.

\section{Conclusions}
In summary, we have demonstrated that magnetic vortex lattices arise under rather general conditions near the magnetic-field-induced quantum critical point that separates magnetically ordered and paramagnetic phases. 
While we have mainly discussed the case of a quantum spin-dimer compound, our theory also applies to nondimerized frustrated magnets near their saturation field $H = H_\text{sat}$. The emergence of magnetic vortex crystals from $6$-${\bf Q}$ condensates in the triangular lattice systems can be immediately generalized to other $p$-fold symmetric lattices with $p \geq 3$. For instance, 4-${\bf Q}$ condensates lead to square crystals of magnetic vortices, and 3-${\bf Q}$ condensates lead to honeycomb crystals, which can be realized in the stacked frustrated square and honeycomb lattices, respectively.

While our theory is only valid in the dilute limit (close to the critical fields), it is interesting to analyze the evolution of the $6$-${\bf Q}$ condensate as a function of increasing density of bosons because this phase could remain stable for higher boson densities. To understand this evolution, we extend our analysis on the nondimerized frustrated magnets near their saturation field. We find that as a function of increasing bosonic density, 
the solid angle wrapped by the spin configuration inside a certain radius from the vortex core should increase [see Fig.~\ref{fig:6-QIIPi}(d)].
Then, there is a critical density, $\rho=\rho_c$, for which the whole sphere (solid angle $4\pi$) is wrapped inside the threshold radius. This means that a skyrmion structure emerges inside of each vortex structure at $\rho=\rho_c$. 
The same solid angle is unwrapped by the spin configuration between the threshold radius and the vortex edge.

The magnetic vortex crystals complement the already-known skyrmion and domain-wall lattices that arise in other families of frustrated magnets. Among other things, the possibility of stabilizing magnetic crystals of topological spin structures opens a new road for studying and exploiting magnetoelectric coupling in Mott insulators.
Magnetic orderings that break enough lattice symmetries induce a redistribution of ionic and electric charge that can lead to a net electric polarization.~\cite{Cheong2007Multiferroics,Bulaevskii2008electronic,Kamiya2012Multiferroic} This magnetically driven ferroelectricity is observed in type-II multiferroic materials.~\cite{Cheong2007Multiferroics}   Because vortex crystals locally break most of the underlying lattice symmetries, they are expected to induce a local charge modulation that becomes more pronounced near the rapidly varying regions of the magnetic configurations (domain walls of soliton lattices and cores of skyrmion and Abelian vortex lattices). Consequently, crystals of topological defects induce charge-density-wave patterns that can be used either to detect these magnetic crystals via x-rays or to achieve magnetic-field control of the local electric polarization and electric-field control of the local spin chirality \cite{Seki2012Observation}.

To conclude, we summarize the essential ingredients for finding magnetic vortex crystals in real materials. The lattice must be $p$-fold symmetric ($p \geq 3$) to allow for a single-boson dispersion with more than two minima. Magnetic frustration is required to have minima at low-symmetry wave vectors. Finally, symmetric exchange anisotropy $J_{A,6}$, which can be produced by either dipolar or spin-orbit interactions, can stabilize this phase over a window of magnetization values that is roughly proportional to $J_{A,6}/J$, where $J$ is the typical value of the exchange coupling. 

\begin{acknowledgments}
  We thank A.~V.~Chubukov, S.~Brown, T.~Okubo, N.~Hatano, T.~Momoi, and G.~Marmorini for valuable discussions.
  Work at LANL was performed under the auspices of the
  U.S.\ DOE Contract No.~DE-AC52-06NA25396 through the LDRD program.
\end{acknowledgments}

\appendix
\section{
  \label{app:ladder}%
  Ladder diagram
}
The Bethe-Salpeter equation for a vertex $\Gamma_{\mathbf{q}}(\mathbf{k},\mathbf{k}')$ defined by the ladder diagram with zero total frequency is
\begin{align} 
  \Gamma_{\mathbf{q}} \left(\mathbf{k},\mathbf{k}'\right)
  = U + V_{\mathbf{q}}
  - 
  \int \frac{\mathrm{d}^3 q'}{8\pi^3}
  \frac{\Gamma_{\mathbf{q}'}\left(\mathbf{k},\mathbf{k}'\right)\,\left(U + V_{\mathbf{q}-\mathbf{q}'}\right)}
       {\omega_{\mathbf{k}+\mathbf{q}'} + \omega_{\mathbf{k}' - \mathbf{q}'}}.
       \label{eq:bethe-salpeter}
\end{align} 
We use a simplified notation $\Gamma_\mathbf{q} \equiv \Gamma_\mathbf{q}\left(\mathbf{k}, \mathbf{k}'\right)$ because the equation does not mix with a different set of $\left(\mathbf{k}, \mathbf{k}'\right)$.
Below we summarize a standard procedure to solve Eq.~\eqref{eq:bethe-salpeter} for lattice models.
We assume the following ansatz:
\begin{align}
  \Gamma_{\mathbf{q}}
  = 
  \left\langle\Gamma\right\rangle
  + \sum_{\eta=1}^{z} A_{\eta}\, V(\mathbf{r}_\eta)\, e^{i\mathbf{q}\cdot\mathbf{r}_\eta},~~
  \left\langle\Gamma\right\rangle
  \equiv 
  \int \frac{\mathrm{d}^3 q'}{8\pi^3}
  \Gamma_{\mathbf{q}'}.
  \label{eq:bethe-salpeter:ansatz}
\end{align}
Here $A_\eta \equiv A_{\eta}(\mathbf{k},\mathbf{k}')$ are undetermined coefficients independent of $\mathbf{q}$.
$\mathbf{r}_\eta$ denotes a displacement vector to a site where the interaction potential due to a particle at the origin is nonzero. $z$ is the total number of such sites.

We assume that the interaction potential is centrosymmetric, meaning $\int \mathrm{d}^3 q'\,V_{\mathbf{q}'} = 0$.
Then, we find
\begin{align}
  \left\langle\Gamma\right\rangle
  =  U \left(
  1 - 
  \int \frac{\mathrm{d}^3 q'}{8\pi^3}
  \frac{\Gamma_{\mathbf{q}'}}{\omega_{\mathbf{k} + \mathbf{q}'}+\omega_{\mathbf{k}' - \mathbf{q}'}}
  \right).
  \label{eq:average-term}
\end{align}
By using this, we can rewrite Eq.~\eqref{eq:bethe-salpeter} as
\begin{align}
  \Gamma_{\mathbf{q}} 
  = \left\langle\Gamma\right\rangle
  + V_{\mathbf{q}}
  - \int \frac{\mathrm{d}^3 q'}{8\pi^3}
  \frac{\Gamma_{\mathbf{q}'}\,V_{\mathbf{q}-\mathbf{q}'}}{\omega_{\mathbf{k} + \mathbf{q}'}+\omega_{\mathbf{k}' - \mathbf{q}'}}.
  \label{eq:bethe-salpeter-form2}
\end{align}
We introduce the following notations:
\begin{align}
  \tau_{0}^{} 
  &\equiv 
  \int \frac{\mathrm{d}^3 q'}{8\pi^3}
  \frac{1}{\omega_{\mathbf{k} + \mathbf{q}'} + \omega_{\mathbf{k}' - \mathbf{q}'}},
  \notag\\
  \tau_{1}^{\eta}
  &\equiv 
  \int \frac{\mathrm{d}^3 q'}{8\pi^3}
  \frac{e^{-i\mathbf{q}'\cdot\mathbf{r}_\eta}}{\omega_{\mathbf{k} + \mathbf{q}'} + \omega_{\mathbf{k}' - \mathbf{q}'}},
  \notag\\
  \tau_{2}^{\eta, \nu}
  &\equiv
  \int \frac{\mathrm{d}^3 q'}{8\pi^3}
  \frac{e^{-i\mathbf{q}'\cdot\left(\mathbf{r}_\eta - \mathbf{r}_{\nu}\right)}}{\omega_{\mathbf{k} + \mathbf{q}'} + \omega_{\mathbf{k}' - \mathbf{q}'}}.
\end{align}
By substituting \eqref{eq:bethe-salpeter:ansatz} into \eqref{eq:average-term}, we obtain
\begin{align}
  \sum_{\eta=1}^{z} V(\mathbf{r}_\eta) \left(\tau_1^\eta\right)^\ast A_\eta + \left(\tau_0^{} + U^{-1}\right) \left\langle\Gamma\right\rangle = 1,
  \label{eq:bethe-salpeter-form3}
\end{align}
where we can take the $U \to \infty$ limit.
In addition, the substitution of \eqref{eq:bethe-salpeter:ansatz} into \eqref{eq:bethe-salpeter-form2} leads to
\begin{align}
  \sum_{\nu=1}^{z} \left(
  \tau_{2}^{\eta,\nu} V(\mathbf{r}_\nu)
  + \delta_{\eta,\nu}
  \right)
  A_\nu 
  + \tau_{1}^{\eta} \left\langle\Gamma\right\rangle
  = 1,~~1 \le \eta \le z.
  \label{eq:bethe-salpeter-form4}
\end{align}
Equations~\eqref{eq:bethe-salpeter-form3} and \eqref{eq:bethe-salpeter-form4} can be made compact 
by introducing $B_{\eta\nu} \equiv \tau_2^{\eta\nu} V(\mathbf{r}_\nu) + \delta_{\eta,\nu}$ and $c_{\eta} \equiv V(\mathbf{r}_\eta)\, \bigl(\tau_1^\eta\bigr)^\ast$:
\begin{align}
  \left(
  \begin{array}{cccc}
    B_{11}&\hdots&B_{1z}&\tau_1^1\\
    \vdots&\ddots&\vdots&\vdots\\
    B_{z1}&\hdots&B_{zz}&\tau_1^z\\
    c_1 & \hdots & c_z & \tau_0^{}
  \end{array}
  \right)
  \left(
  \begin{array}{c}
    A_1\\
    \vdots\\
    A_z\\
    \left\langle\Gamma\right\rangle
  \end{array}
  \right)
  =
  \left(
  \begin{array}{c}
    1 \\
    \vdots\\
    1\\
    1
  \end{array}
  \right).
  \label{eq:linear-eq}
\end{align}
Because the matrix elements on the left-hand side are $\mathbf{q}$ independent, $A_\eta$ ($1 \le \eta \le z$) and $\langle\Gamma\rangle$ are also $\mathbf{q}$ independent, as is assumed in the ansatz~\eqref{eq:bethe-salpeter:ansatz}. 
While lattice symmetries may be used to simplify Eq.~\eqref{eq:linear-eq}, this is a generic prescription for solving the integral equation~\eqref{eq:bethe-salpeter} for lattice models.

\section{
  \label{app:spinconfig}%
  Spin configuration for a given BEC state
}
In the BEC phase, the state of an individual dimer (for a dimerized magnet at $H \sim H_{c}$) or a spin (for $H \sim H_\text{sat}$) at $\mathbf{r}$ is well approximated by the coherent state:\cite{Kolezhuk1996Continuum-field}
\begin{align}
  \left\lvert{\psi}\right\rangle_\mathbf{r}
  \approx 
  \sqrt{1 - \left\lvert{\psi_\mathbf{r}}\right\rvert^2} \lvert{\widetilde{\uparrow}}\rangle_\mathbf{r}
  + \psi_\mathbf{r} \lvert{\widetilde{\downarrow}}\rangle_\mathbf{r},
\end{align}
where
\begin{align}
  \psi_\mathbf{r} 
  = \sum_{n=1}^{3}
  \left(
  \rho^{}_{\mathbf{Q}_n} e^{i\phi^{}_{\mathbf{Q}_n}} e^{i\mathbf{Q}_n \cdot \mathbf{r}}
  + \rho^{}_{-\mathbf{Q}_n} e^{i\phi^{}_{-\mathbf{Q}_n}} e^{-i\mathbf{Q}_n \cdot \mathbf{r}}
  \right).
\end{align}
The expectation value of the pseudospin $\widetilde{\mathbf{S}}_\mathbf{r}$ is
\begin{align}
  \left\langle{\widetilde{S}^+_\mathbf{r}}\right\rangle 
  &= \sqrt{1 - \left\lvert{\psi_\mathbf{r}}\right\rvert^2}\; \psi_\mathbf{r},
  \notag\\
  \left\langle{\widetilde{S}^z_\mathbf{r}}\right\rangle 
  &= \frac{1}{2} - \left\lvert{\psi_\mathbf{r}}\right\rvert^2,
\end{align}
which implies $0 \le \left\lvert{\psi_\mathbf{r}}\right\rvert \le 1$.
$\bigl\langle{\widetilde{S}^x_\mathbf{r}}\bigr\rangle^2 + \bigl\langle{\widetilde{S}^y_\mathbf{r}}\bigr\rangle^2 + \bigl\langle{\widetilde{S}^z_\mathbf{r}}\bigr\rangle^2 = 1/4$ is guaranteed.
This pseudospin is a real spin for a spin-$1/2$ system close to the saturation field.

To obtain the spin configuration for a dimerized magnet near the field-induced QCP, where we regard $\lvert{\widetilde{\uparrow}}\rangle_\mathbf{r}$ and $\lvert{\widetilde{\downarrow}}\rangle_\mathbf{r}$ as the singlet and the triplet polarized along the external-field direction, respectively, we evaluate the expectation values of the spin operators of each spin, $a=1,2$, on a dimer at $\mathbf{r}$.
For instance, the expressions for $S=1$  are
\begin{align}
  \left\langle{S^+_{\mathbf{r},a}}\right\rangle 
  &= \left(-1\right)^a \frac{2}{\sqrt{3}}
  \sqrt{1 - \left\lvert{\psi_\mathbf{r}}\right\rvert^2}\; \psi_\mathbf{r}^{\ast},
  \notag\\
  \left\langle{S^z_{\mathbf{r},a}}\right\rangle 
  &= \frac{1}{2} \left\lvert{\psi_\mathbf{r}}\right\rvert^2.
\end{align}

\bibliography{references}
\end{document}